\documentclass[10pt,a4]{article}

\usepackage[pdftex,colorlinks=true,citecolor=blue,urlcolor=blue,linkcolor=black]{hyperref}
\usepackage{graphicx}
\usepackage[pdftex]{color}

\usepackage{stix}
\usepackage{amsmath}
\usepackage{amssymb}
\usepackage{cite}

\def\lambdabar{\,{\raisebox{0.45ex}{-}\mkern-10mu\lambda}}

\newcommand{\citelow}[1]{\cite{#1}}
\let\oldbibitem\bibitem%
\renewcommand{\bibitem}[1][1]{\oldbibitem}

\begin{document}

\markboth{Carsten Henkel}
{Electromagnetic Response of Metals}

\title{\protect\textbf{Electromagnetic Response of the Electron Gas\\
and the Thermal Casimir Pressure Anomaly}}
\author{%
Carsten Henkel
\\[1ex]
\it Institute of Physics and Astronomy, University of Potsdam,
\\
\it Karl-Liebknecht-Str. 24/25,
14476 Potsdam, Germany}

\date{2024 Dec 10}

\maketitle

\begin{abstract}
A review of the nonlocal electromagnetic response functions 
for the degenerate electron gas, computed within standard perturbation
theory, is given. 
These expressions due to Lindhard, Klimontovich and Silin 
are used to re-analyze the Casimir interaction 
between two thick conducting plates 
in the leading order at high temperatures
(zero'th term of Matsubara series).
Up to small corrections that we discuss, the results of the conventional
Drude model are confirmed.
The difference between longitudinal and transverse permittivities 
(or polarization tensors) yields the Landau (orbital) diamagnetism of 
the electron gas.
\end{abstract}

\section{Introduction}

One of the driving forces behind measurements of Casimir
forces goes beyond this fundamental interaction, namely to establish
precision data for gravitational interactions on the scale of a few
microns and smaller \cite{Decca_2005}.
Progress in that direction is hampered because unfortunately, the
dispersion force currently cannot be subtracted simply as a
systematic effect. 
This is related to its anomalously strong temperature dependence 
between electrical conductors. 
The comparison with the case of ideally reflecting 
materials \cite{Mehra_1967} reveals that deviations from the pure
quantum setting (temperature $T = 0$) appear at unusually small
length scales, much shorter than the Wien wavelength $\lambda_T
= \hbar c / k_B T \approx 7.8\,\mu{\rm m}$ (at room temperature)
\cite{Bostroem_2000}. 
This thermal anomaly is due to excitations
in the conducting walls with a dispersion relation distinct from
light in vacuum.
If we introduce 
the lowest cavity mode between ideally reflecting walls at distance $d$ as
a characteristic energy scale 
$E_{\rm id} = \hbar c / 2d$,
thermal effects are expected to appear
for $k_B T \gg E_{\rm id}$, i.e.\ for 
$d \gg \lambda_T/2$, as confirmed by Mehra's calculation
\cite{Mehra_1967},
see also Fig.\,1 in Ref.\citelow{Klimchitskaya_2022d}.
But in a normal conductor,
the diffusion of magnetic fields across the characteristic scale $d$,
with a diffusion constant $D_m = 1/\mu_0 \sigma$ set by the DC conductivity,
gives the energy scale 
$E_{\rm diff} = \hbar D_m / d^2$, analogous to the Thouless energy.
This moves the region of temperature-dependent physics 
down to much smaller distances, 
$d \gg \lambdabar_p (\hbar / k_B T \tau)^{1/2}$ 
where $\lambdabar_p = c / \Omega$
is the reduced plasma wavelength 
($\approx 22\,{\rm nm}$ for gold),
$\Omega$ the plasma frequency, 
and $\tau$ the Drude scattering time 
($\hbar / k_B T \tau \approx 0.95$ at room temperature)
\cite{Intravaia_2009a}.

The comparison to experimental data \cite{Decca_2005, Banishev_2013b, 
Liu_2019a} has narrowed down the discussion to the evaluation of 
the zero'th term in the Matsubara representation of the Casimir
energy. 
This term is of classical origin \cite{Feinberg_2001} except
for quantum physics being possibly hidden in the material properties.
An agreement with experiment is obtained when the permittivities
of the metals are modelled according to the so-called plasma model
(that describes, at low frequencies, the electromagnetic response
of superconductors \cite{London_1935, Steinberg_2008}).
The physically more intuitive Drude model with a finite DC conductivity
gives predictions that deviate between 10\% up to a factor $10^3$ 
from the observations \cite{Klimchitskaya_2022d}.
This discrepancy is the motivation for the present paper.

We review the model put forward by Lindhard \cite{Lindhard_1954}
and Klimontovich and Silin \cite{Klimontovich_1960}
for the dielectric functions of the electron gas 
(Sec.\,\ref{s:Lindhard-basics}, Appendix~\ref{a:Lindhard-functions}).
We use these results to evaluate the zero'th order Matsubara term,
in particular the reflection amplitudes for quasi-static electric
and magnetic fields. 
The upshot is that these are consistent with
the predictions of the Drude model, with a small correction due
to Landau (orbital) diamagnetism \cite{AshcroftMermin_book}. 
The inclusion of a finite carrier lifetime introduces some quantitative
differences, but the main conclusions are robust.

Before starting the technical discussion, a brief remark about
dissipative media and Casimir physics. Rytov's formulation of
fluctuation-induced forces seems to indicate that the electromagnetic
stress tensor vanishes when material losses (the imaginary part of
the dielectric function) are put to zero. This conclusion is, however, 
short-sighted. Indeed, in this limit, the source currents in the
dissipative material actually morph into the field mode operators 
of the given geometry and, so to speak, lose their ``matter character''.
The field modes indeed become well-defined
with real-valued frequencies, as the limit is performed
\cite{Scheel_1998a, Savasta_2001}.

Another delicate issue with losses can be illustrated with 
Lindhard's dielectric functions recalled below: even if the carriers
have well-defined states (i.e., real-valued excitation 
energies, no collisions), there is a nonzero imaginary part to the
dielectric functions. It translates the excitation of electron-hole
pairs by electromagnetic fields (Landau damping)
\cite{DresselGruener_book}.
The region in the $q\omega$-plane where this happens 
(\emph{cf.}\ Fig.\,\ref{fig:absorption} in Appendix~\ref{a:Lindhard-functions})
is relatively robust with respect to the introduction of 
a finite carrier lifetime $\tau$. 
The latter mostly affects the region of small frequencies
$\omega < 1/\tau$ (at room temperature, wavelengths longer 
than $\sim 50\,\mu{\rm m}$). 
A consistent treatment of carrier collisions has to address 
conservation laws like particle number and potentially the total
momentum (for electron-electron processes, for example). In the
literature, this has been discussed in several places,
an incomplete list of references 
includes Refs.\citelow{Warren_1960, Mermin_1970, Liu_1970, Ford_1984,
Conti_1999, Roepke_1999}.
This may provide some guiding to the analogous problem in
graphene whose conductivity has been subject to some discussion
in the recent literature
\cite{Khusnutdinov_2024, RodriguezLopez_2024a}.

\section{Static limit of reflection problem}

The thermal anomaly of the Casimir pressure between metals arises,
as discussed in the Introduction, from the zero'th term in the
Matsubara series. 
This involves the reflection of electromagnetic waves 
in the limit of zero frequency, 
and we re-consider this problem here.

\subsection{Conventional local approximation}

\begin{figure}[htbp]
   \centering
   \includegraphics[height=0.42\textwidth]{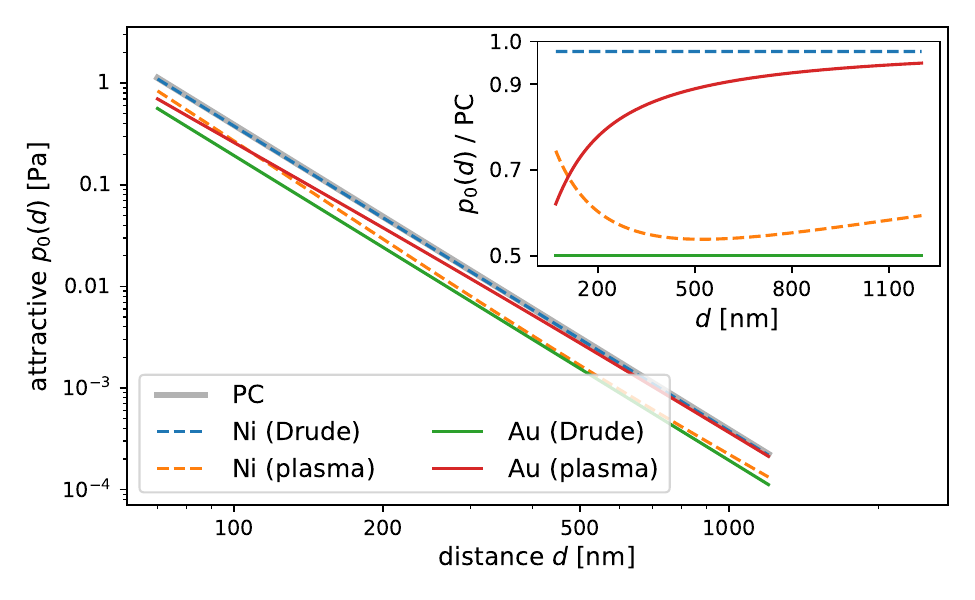}
   \caption[]{Casimir pressure Eq.\,(\ref{eq:zero-Mb-pressure}) 
   between planar surfaces
   in the leading order at high temperatures
   (zero'th term of Matsubara series).
   The Drude and plasma models are defined by Eqs.\,(\ref{eq:rp-limit},
   \ref{eq:rs-plasma-Drude}), respectively. Parameters:
   $T = 293\,{\rm K}$, Gold: $\hbar\Omega = 9.1\,{\rm eV}$,
   Nickel: $\hbar\Omega = 4.9\,{\rm eV}$ 
   and $\mu = 110\,\mu_0$. PC: perfectly reflecting surfaces.
   }
   \label{fig:pressure_Mb0}
\end{figure}

When the Fresnel formulas are used, they require the 
dielectric function $\varepsilon(\omega)$. We take incident fields with a
wave vector $K$ parallel to the surface (along the $x$-direction, say).
In the medium, the fields behave exponentially with the decay constant
$\kappa_m$ given by $\kappa_m^2 = K^2 - \omega^2 \varepsilon(\omega) \mu$.
While the permittivity $\varepsilon(\omega)$ diverges in the static limit for a conducting medium, the limit of $\kappa_m^2$ depends on the model. 
In the so-called plasma model, $\varepsilon(\omega)$ has
a second-order pole and
\begin{equation}
\text{plasma model:} \qquad
\kappa_m \to \sqrt{ K^2 + \Omega^2 \varepsilon_0\mu }
\,,
\label{eq:}
\end{equation}
where $\mu$ is the static permeability.
In the Drude model, $\omega \to 0$ reveals a pole of first order and
\begin{equation}
\text{Drude model:} \qquad
\kappa_m \to K
\,,
\label{eq:}
\end{equation}
independently of any other parameters.
The resulting static Fresnel coefficients are, using 
$k^2 = (\omega/c)^2 - K^2$ on the vacuum side,
\begin{equation}
r_p = \frac{ {\rm i} \varepsilon_0 \kappa_m - \varepsilon( \omega ) k
}{ {\rm i} \varepsilon_0 \kappa_m + \varepsilon( \omega ) k }
\to 
-1
\label{eq:rp-limit}
\end{equation}
in the p- or TM-polarization. 
This applies to both models. A difference occurs in the s- or TE-polarization
\begin{equation}
r_s = \frac{ \mu k - {\rm i} \mu_0 \kappa_m }{ \mu k +  {\rm i} \mu_0 \kappa_m }
\to 
\begin{cases}
\displaystyle
\frac{ \mu K - \mu_0 \sqrt{ K^2 + \Omega^2 \varepsilon_0\mu } 
}{ \mu K + \mu_0 \sqrt{ K^2 + \Omega^2 \varepsilon_0\mu } }
& 
\text{(plasma)}
\\[3.5ex]
\displaystyle
\frac{ \mu - \mu_0 }{ \mu + \mu_0 }
& 
\text{(Drude)}
\end{cases}
\label{eq:rs-plasma-Drude}
\end{equation}
Normal metals like gold are considered non-magnetic so that $\mu = \mu_0$,
and $r_s \to 0$ in the Drude model.
This is the usually considered plasma-Drude difference.
Indeed, the Casimir pressure arising from the zero'th term in the Matsubara
series is (positive sign for attraction)
\cite{Klimchitskaya_2022d}
\begin{equation}
p_{0}(d) = k_B T \!\int\limits_0^\infty\!\frac{{\rm d}K}{2\pi} 
\, K^2 \sum_{\sigma \,=\, s,\,p} 
\frac{ r_\sigma^2(K, 0) \, {\rm e}^{ - 2 K d } } { 1 - r_\sigma^2(K, 0) \, {\rm e}^{ - 2 K d } }
\label{eq:zero-Mb-pressure}
\end{equation}
with results plotted in Fig.\,\ref{fig:pressure_Mb0}.  
The inset compares the data to a perfect reflector defined as
$|r_s| = |r_p| = 1$ where $p_0(d) = (\zeta(3)/4\pi)\, k_B T / d^3$.
The Drude model for a magnetic material like nickel
nearly doubles the thermal pressure relative 
to a non-magnetic one, due to the large value of $\mu/\mu_0$. 
In the plasma model, both cases lead to fractional power laws (see inset)
because the reflection coefficient $r_s$ is $K$-dependent.
Experimental data are found to agree with the plasma model (when adding
all terms of the Matsubara series). 
As illustrated in Figs.\,3 and~7 of Ref.\citelow{Klimchitskaya_2022a},
the Drude model underestimates the Casimir pressure between gold bodies,
while it overestimates it between nickel bodies.
The mixed gold/nickel case is well described by the Drude model
(Fig.\,6 of Ref.\citelow{Klimchitskaya_2022a}).

\subsection{Lindhard dielectric functions}
\label{s:Lindhard-basics}

In the following, we re-analyze the zero-frequency Matsubara term for
the Casimir pressure, starting from the Lindhard dielectric functions. 
It is worth recalling that these formulas are
based on first principles, namely the response of the ideal Fermi
gas to an electromagnetic field.
At this stage of the model, except the standard perturbation theory,
no further approximations are made; in particular, all electronic states have
well-defined energies.
The issue of carrier scattering, which is crucial for any realistic material, 
will be addressed in Sec.\,\ref{s:with-losses}.

We outline in Appendix~\ref{a:spatial-dispersion} the concept of spatial
dispersion. 
The permittivity then depends on both frequency and wavelength 
of the electromagnetic field, and it becomes a tensor.
Two functions $\varepsilon_L(q, \omega)$, $\varepsilon_T(q, \omega)$
are introduced that provide the response
to longitudinal (${\bf E}_L = - \nabla \phi$) and transverse fields
(${\bf E}_T = 
{\rm i} \omega {\bf A}$ with $\nabla \cdot {\bf A} = 0$).
Lindhard gives for a fully degenerate Fermi gas
(temperature $k_B T \ll E_F$, Fermi energy) the following
parametrisation of the longitudinal dielectric function
\cite{Lindhard_1954} 
\begin{align}
\frac{ \varepsilon_L( q, \omega ) }{ \varepsilon_0 } &=
1 + \frac{ 3 \Omega^2 }{ v_F^2 q^2 } f_L( z, u )
\label{eq:epsL-Lindhard}
\\
f_L( z, u ) &=
\frac12 
+
\frac{ 1 - (z - u)^2 }{ 8 z }
\log \frac{ z - u + 1 }{ z - u - 1 }
+
\frac{ 1 - (z + u)^2 }{ 8 z }
\log \frac{ z + u + 1 }{ z + u - 1 }
\label{eq:def-fL-q}
\end{align}
Here, the response of bound electrons and ionic cores is neglected,
$\Omega$ is the plasma frequency, 
and the conventional Lindhard variables are used
\begin{equation}
u = \frac{ \omega + {\rm i} 0 }{ v_F q }
\,,\qquad
z = \frac{ q }{ 2 k_F }
\label{eq:}
\end{equation}
where $m v_F = \hbar k_F$ is the Fermi momentum.
For the $+ {\rm i} 0$ prescription, see Sec.\,\ref{s:with-losses}.
The real and imaginary parts of the Lindhard function arise from a
careful analysis of the branch points of the logarithm (to basis $e$) 
and are given in 
Appendix~\ref{a:Lindhard-functions}. %
The transverse dielectric function found by Lindhard is defined by
\begin{align}
\frac{ \varepsilon_T( q, \omega ) }{ \varepsilon_0 } &=
1 - \frac{ \Omega^2 }{ \omega^2 } f_T(z, u) 
\label{eq:epsT}
\\
f_T( z, u ) &=
\frac{ 3 }{ 8 }( 1 + 3 u^2 + z^2 )
\nonumber\\
& \phantom{=
}
- 
\frac{ 3 [1 - (z-u)^2]^2}{ 32 z} \log \frac{ z - u + 1 }{ z - u - 1 }
- 
\frac{ 3 [1 - (z+u)^2]^2}{ 32 z} \log \frac{ z + u + 1 }{ z + u - 1 }
\label{eq:def-fT-q}
\end{align}

Nonlocal dielectric functions have been used in earlier work 
to compute the Casimir pressure at zero temperature, as well as the
associated entropy \cite{Esquivel_2004a, ContrerasReyes_2005a, Sernelius_2005a, Svetovoy_2006a, Dalvit_2008, Pitaevskii_2008, Svetovoy_2008b}.
Deviations from a local dielectric function are largest
at distances of the order of $\lambdabar_p$ or smaller,
but are at most $\approx 0.5\%$ for $d > 60\,{\rm nm}$.%

\subsection{Low-frequency limit of reflectivities}

In analyzing the low-frequency regime,
our approach is opposite to the small-$q$ limit
that was considered in Ref.\citelow{Hannemann_2021}
and where the Lindhard functions were expanded for $u \to \infty$ 
and $z \to 0$.
Here, we keep $q \sim 1/d$ and take first the limit $u \to 0$.

Performing the expansion 
$u \to 0$ and keeping only a few low terms,
\begin{equation}
\begin{aligned}
\frac{
\varepsilon_L( q, \omega \to 0 )
}{ 
\varepsilon_0 }
 &=
1 + \frac{ 3 \Omega^2 }{ q^2 v_F^2 } + 
\frac{ 3\pi {\rm i} \Omega^2 \omega }{ 2 q^3 v_F^3 }
\,,\quad &
\frac{ 
\varepsilon_T( q, \omega \to 0 )
}{ 
\varepsilon_0 }
 &=
1 - \frac{ \Omega^2 q^2 }{ 4 \omega^2 k_F^2 }
+ \frac{ 3\pi {\rm i} \Omega^2 }{ 4 \omega q v_F }
\end{aligned}
\label{eq:static-eps-LT}
\end{equation}
The last terms arise from the branch points of the logarithms 
in Eqs.\,(\ref{eq:def-fL-q}, \ref{eq:def-fT-q}).
Note that the leading terms do not resemble the plasma model, 
and that a significant $q$-dependence (spatial dispersion)
appears in both cases. 

We now address the reflection and transmission problem at the interface
between vacuum and the Lindhard gas. This typically requires matching
rules for the material excitations, also known as ``additional
boundary conditions'' 
\cite{Forstmann_1978, Henneberger_1998, Silveirinha_2009}.
Their construction in the following is relatively straightforward
if we focus on the static limit.
The idea is to re-write the material equations as differential equations.

\subsubsection{Longitudinal fields, p-polarisation}

By the definition of the dielectric constant, the longitudinal polarization field
in the medium is given by 
${\bf P}_L = \big( \varepsilon_L - \varepsilon_0 \big) \, {\bf E}_L$
where ${\bf E}_L = -\nabla \phi$ is the longitudinal electric field.
Taking the leading order of Eq.\,(\ref{eq:static-eps-LT}) and 
re-writing the $q^2$ factor as spatial derivatives, we get
\begin{equation}
- \nabla^2 {\bf P}_L = \frac{ 3 \Omega^2 }{ v_F^2 } \varepsilon_0 (-\nabla \phi)
\,,\qquad
\nabla^2 \rho = \frac{ 3 \Omega^2 }{ v_F^2 } \rho = \frac{ \rho }{ \Lambda^2 }
\,.
\label{eq:Thomas-Fermi-bulk}
\end{equation}
The second equality is obtained by taking the divergence on both sides;
it embodies the screening
of the charge density \cite{Warren_1960}
on the Thomas-Fermi scale $\Lambda = v_F / (\sqrt{3}\, \Omega)$.

Considering a planar interface between two half-spaces, 
all fields may be taken proportional to 
${\rm e}^{ {\rm i} K x }$.
This allows for an exponentially decaying solution
\begin{equation}
z \ge 0: \quad
\rho(z) = \rho(0) \, {\rm e}^{-\kappa z}
\qquad \text{with} \qquad
\kappa^2 = K^2 + 1 / \Lambda^2
\label{eq:charge-density}
\end{equation}
that describes the charge induced by the incident field.
The basic equations to solve next to Eq.\,(\ref{eq:Thomas-Fermi-bulk}) are now
\begin{align}
- \nabla^2 \phi &= \rho / \varepsilon_0
\,,\qquad &
- \nabla \cdot {\bf P}_L &= \rho
\label{eq:}
\end{align}
with sources localized in the material. On the vacuum side, we consider
a combination of ``incident'' and ``reflected'' fields
\begin{equation}
z \le 0: \quad 
\phi(z) = {\rm e}^{ - K z } + r_p \, {\rm e}^{ K z }
\label{eq:}
\end{equation}
which would be produced by a charge distribution periodic along the $x$-direction
and localised in $z < 0$.
The corresponding electric field vector ${\bf E}_L$ lies in the $xz$-plane,
as expected for the p-polarization. 
In the metal, the potential has two terms
\begin{equation}
z \ge 0: \quad 
\phi(z) = t_p \, {\rm e}^{ - K z } + t_L \, {\rm e}^{ -\kappa z }
\,,
\label{eq:}
\end{equation}
the second one being due to the charge density~(\ref{eq:charge-density}).
A similar \emph{Ansatz} is made for the polarization field ${\bf P}_L$.

For a complete specification of this interface problem, 
let us assume the potential $\phi$
and its derivative ${\bf E}_L$ to be continuous across the interface. This
excludes $\delta$-like surface charges, a natural condition, since the spatial 
extent ${\cal O}(\Lambda)$ of the charge distribution is actually included 
in the model.
The polarization ${\bf P}_L$ is an auxiliary quantity introduced to 
represent the medium charge. 
Here, we impose it to vanish at the surface,
${\bf P}_L(0) = 0$, ensuring continuity with the vacuum half-space.
In a time-dependent setting where the current density is ${\bf j}_L = 
- {\rm i} \omega {\bf P}_L$, these boundary conditions would ensure charge
conservation (no outflow of the metal) and a no-slip tangential current.

\begin{figure}[bthp]
   \centering
   \includegraphics[height=0.37\textwidth]{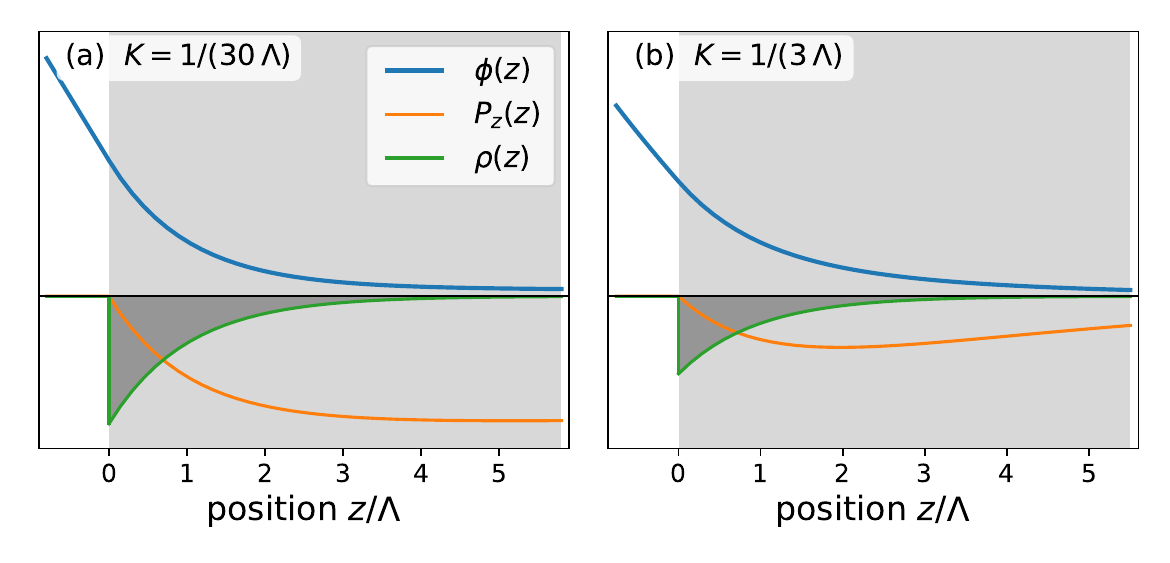} %
   \caption[]{Static electric field applied to an electron gas half-space
   described by the longitudinal Lindhard function $\varepsilon_L(q, \omega \to 0)$
   of Eq.\,(\ref{eq:static-eps-LT}). Electric potential, polarization field
   and charge density have been scaled to units common to both panels.
   The electric field (pointing into the metal) creates an exponential
   charge depletion zone below the surface (dark shaded area).
   $\Lambda$ is the Thomas-Fermi screening length of 
   Eq.\,(\ref{eq:Thomas-Fermi-bulk}).
   }
   \label{fig:static-rp-solution}
\end{figure}

The solution to this problem is illustrated in 
Fig.\,\ref{fig:static-rp-solution} for two values of the ratio
$\kappa/K$. The reflection coefficient and the amplitude of the sub-surface
charge are given by
\begin{equation}
r_p(K, 0) = \frac{ - \kappa^2 + K^2 }{ \kappa^2 + 2 \kappa K + 5 K^2 }
\,,\qquad
\rho(0) = \frac{ 6 \varepsilon_0 \Omega^2 }{ v_F^2 }
\frac{ K (K + \kappa) }{ \kappa^2 + 2 \kappa K + 5 K^2 }
\label{eq:static-rp}
\end{equation}
where $\rho(0)$ is measured relative to a unit incident potential.
Equation~(\ref{eq:static-rp}) replaces the Fresnel result 
$r_p = -1$ of Eq.\,(\ref{eq:rp-limit}),
which is recovered, however,
for $K \ll \kappa$, i.e., for wavelengths much longer than the
Thomas-Fermi screening length. 
Because the latter is very small ($\Lambda\approx 0.1\,{\rm nm}$
for gold), the reflection coefficient~(\ref{eq:static-rp}) does not generate
any significant difference compared to the common choice $r_p = -1$.
The result $r_p(K, 0) = -(1 - K \Lambda)/(1 + K \Lambda)$ obtained 
in Ref.\citelow{Svetovoy_2005a} within a nonlocal approach
is also consistent with Eq.\,(\ref{eq:static-rp}) to first order
in $K \Lambda$.

On length scales as short as $\Lambda$, one expects, of course,
additional phenomena to become relevant like the smooth equilibrium
charge profile (showing some spill-out relative to the nominal surface
plane $z = 0$), the crystalline lattice of the positive ions and their
polarization.
These issues have revived attention quite recently in the field of
nano-scale plasmonics \cite{Yang_2019c, Mortensen_2021b}.
The impact of a smooth charge density on the Casimir pressure
is smaller than $\approx 0.5\%$, however, at distances larger than
$60\,{\rm nm}$, see Ref.\citelow{ContrerasReyes_2005a}.
Short-scale physics may go beyond the basic assumption
behind Lifshitz theory, namely that material parameters do not depend
on the presence of a second body, a caveat formulated by Barash
and Ginzburg \cite{Barash_1975}.

\subsubsection{Transverse fields, s-polarisation}
\label{s:s}

The second relation in Eq.\,(\ref{eq:static-eps-LT}) is related to magnetism. 
Restricting again to the leading order as $\omega \to 0$, 
we have for the transverse polarisation
\emph{current}
${\bf j}_T = - {\rm i} \omega {\bf P}_T$
using $\nabla^2 {\bf E}_T = - \nabla \times (\nabla \times {\bf E})$,
\begin{equation}
{\bf j}_T %
= - \frac{ \Omega^2 \varepsilon_0 }{ 4 {\rm i}\omega k_F^2 } \nabla \times (\nabla \times {\bf E})
= - \frac{ \Omega^2 \varepsilon_0 }{ 4 k_F^2 } \nabla \times {\bf B}
\,.
\label{eq:transverse-current}
\end{equation}
Writing the lhs as a magnetization current ${\bf j}_T = \nabla \times {\bf M}$,
we recognize the static magnetic response
${\bf M} = \chi {\bf H} = \chi {\bf B}/\mu$
with the susceptibility
\begin{equation}
\frac{ \chi }{ \mu } = - \frac{ \Omega^2 \varepsilon_0 }{ 4 k_F^2 } 
= - \frac{ \Omega^2 }{ 4 \mu_0 c^2 k_F^2 }
\,.
\label{eq:Landau-chi}
\end{equation}
This is consistent with the general relation~(\ref{eq:T-L-and-mu}) 
of Appendix~\ref{a:spatial-dispersion}
between
the permeability, on the one hand, 
and the longitudinal and transverse permittivities 
of Eq.\,(\ref{eq:static-eps-LT}), on the other:
\begin{equation}
\lim_{\omega\to0}
\Big(
1 - \frac{ \mu_0 }{ \mu } 
\Big)
=
- \lim_{\omega\to0}
\frac{\omega^2}{c^2 q^2} 
\bigg(
\frac{ \Omega^2 q^2 }{ 4 \omega^2 k_F^2 }
+ 
\frac{ 3 \Omega^2 }{ v_F^2 q^2 } 
\bigg)
= 
- \frac{ \Omega^2 }{ 4 c^2 k_F^2 }
\,.
\label{eq:}
\end{equation}
The susceptibility~(\ref{eq:Landau-chi}) represents the 
Landau diamagnetism of the Fermi gas,
equal to $-\frac13$ of its susceptibility due to the
electron spin (Pauli paramagnetism).
It can also be expressed as
$\chi = - \tfrac{1}{3} \mu_0 \mu_B^2 g(E_F)$ with the density of states
at the Fermi energy $g(E_F) = 3n_0 / (2E_F)$,
$n_0$ the conduction electron density, 
and the Bohr magneton $\mu_B = e \hbar / (2 m)$. 
This formula gives for gold the small value $\chi \approx -3.6\times 10^{-6}$.
Experimental data 
point towards $-3.45 \times 10^{-5}$, while metallic nanostructures 
even show ``giant orbital diamagnetism'' with a susceptibility increasing
up to $-4.9 \times 10^{-4}$, see Refs.\citelow{BleszynskiJayich_2009, vanRhee_2013}.
For estimates of the number of surface states involved in this giant magnetic
response and the mesoscopic fluctuations in the number of energy levels in
a small, disordered metallic grain with significant spin-orbit interaction, 
see Refs.\citelow{Hernando_2014, Murzaliev_2019}.

The reflection problem is easier to solve in the case of transverse
fields, as the material develops effectively a local magnetic response
described by the permeability $\mu$. %
Transverse fields
are described by the vector potential ${\bf A}$ in the Coulomb gauge.
The incident electric field vanishes in the static limit,
${\bf E}_T = {\rm i} \omega {\bf A} \to {\bf 0}$,
so that only incident and reflected magnetic fields remain:
\begin{equation}
z \le 0: \quad
A_y(z) = {\rm e}^{ - K z } + r_s\, {\rm e}^{ K z }
\,.
\label{eq:Ay-incident}
\end{equation}
The $y$-component of the vector potential, perpendicular to the plane
of incidence, is characteristic for the s-polarization.
In the medium, we have
\begin{equation}
\nabla \times {\bf B}/\mu_0 %
= {\bf j}_{T} = \nabla \times \chi {\bf B}/\mu
\label{eq:}
\end{equation}
with the susceptibility of Eq.\,(\ref{eq:Landau-chi}).
This leads to ${\bf H} = {\bf B}/\mu_0 - \chi {\bf B}/\mu
= {\bf B}/\mu$, i.e., 
$\mu = \mu_0 ( 1 + \chi )$,
and finally $\nabla^2 A_y = 0$ again.
The Ampère(-Maxwell) equation is hence solved with the Ansatz
\begin{equation}
z \ge 0: \quad
A_y(z) = t_s\, {\rm e}^{ - K z }
\,.
\label{eq:Ay-in-medium}
\end{equation}
The boundary condition is that $B_z = {\rm i} K A_y$ 
and $H_x = B_x / \mu(z)$ are continuous,
to avoid $\delta$-like magnetic surface charges or currents 
from arising in the equations
$\nabla \cdot {\bf B} = 0$ and $\nabla \times {\bf H} = {\bf 0}$.
The solution is the reflection coefficient
\begin{equation}
r_s(K, 0) 
= \frac{ \mu - \mu_0 }{ \mu + \mu_0 }
\approx \frac{ \chi }{ 2 }
\label{eq:}
\end{equation}
where the last equality takes into account $|\chi| \ll 1$.
This calculation within the Lindhard model for the
electron gas
hence confirms
the Drude choice for the reflection coefficient,
Eq.\,(\ref{eq:rs-plasma-Drude}).
The numerical difference due to the nonzero susceptibility $\chi$ is negligible,
however,
since Landau diamagnetism is so weak, even if it is ``giant''.

We recall that the Lindhard dielectric functions for the electron gas
contain losses because their imaginary part is nonzero. 
This translates the excitation of electron-hole pairs by lifting an
electron from the Fermi ball to a vacant state.   
The finite lifetime of carrier states has not been included yet. 
This is the topic of the following section.

\subsection{Impact of carrier scattering}
\label{s:with-losses}

To set the stage for this problem, we recall the perturbative result 
for the current induced in a homogeneous electron gas
by a transverse electric field
with momentum
$\hbar{\bf q}$ and frequency $\omega$
\cite{Lindhard_1954, Warren_1960}
\begin{equation}
{\bf j} = \frac{ 2 }{ (2\pi\hbar/m)^3 }
\int\!{\rm d}^3v\, 
\frac{ e {\bf v} }{ 
	1/\tau - {\rm i} (\omega + {\bf q} \cdot {\bf v} )
	}
\frac{e {\bf E}_T}{m} \cdot \frac{ \partial f^{(0)} }{ \partial {\bf v} }
\,.
\label{eq:Boltzmann-response}
\end{equation}
For simplicity, this is written down in a semiclassical formulation
where the equilibrium phase space distribution is given by $f^{(0)}({\bf v})$, 
and its perturbation
evolves according to the Boltzmann equation. The results discussed so
far are obtained with an infinitesimal scattering rate $1/\tau = 0^+$ which boils
down to a prescription how to avoid the pole in the 
integral~(\ref{eq:Boltzmann-response}).
In the quantum-mechanical treatment, a similar denominator arises from the
energy balance in the absorption of one energy quantum by an electron 
with momentum $m{\bf v}$:
\begin{equation}
E(m{\bf v} + \hbar{\bf q}) - E(m{\bf v}) - \hbar\omega
=
\hbar {\bf q} \cdot {\bf v} + \frac{ \hbar^2 {\bf q}^2 }{ 2 m } - \hbar \omega
\,.
\label{eq:energy-difference}
\end{equation}
The term proportional to $\hbar^2$ is a ``quantum correction'' to
Eq.\,(\ref{eq:Boltzmann-response}) that was included in 
Eqs.\,(\ref{eq:def-fL-q}, \ref{eq:def-fT-q}) above. 

In realistic materials, the electronic states have a finite lifetime
due to scattering among carriers or off impurities etc. To include
this, Lindhard gives the ``difference of electronic
energies an imaginary part $-{\rm i} \hbar / \tau$,'' making the replacement
\cite{Lindhard_1954}
\begin{equation}
\hbar {\bf q} \cdot {\bf v} + \frac{ \hbar^2 {\bf q}^2 }{ 2 m } - \hbar \omega
\mapsto
\hbar {\bf q} \cdot {\bf v} + \frac{ \hbar^2 {\bf q}^2 }{ 2 m } - \hbar \omega
- {\rm i}\hbar/\tau
\,.
\label{eq:complex-denominator}
\end{equation}
This shifts the $\omega$-pole into the lower half-plane. 
The expression~(\ref{eq:complex-denominator}) can also be interpreted
as a complex frequency $\varpi = \omega + {\rm i}/\tau$. 
For infinite $\tau$, this would mimick the adiabatic switching on of
the perturbation $\sim \exp(- {\rm i} \varpi t) =
{\rm e}^{ - {\rm i}\omega t } \exp(t/\tau)$, a common choice to compute 
the retarded system response.

It has been recognized by Warren and Ferrell \cite{Warren_1960} 
and by Mermin \cite{Mermin_1970} that
the Boltzmann result~(\ref{eq:Boltzmann-response}) cannot be used for a
finite carrier lifetime $\tau$. One has to take into account that 
scattering processes conserve charge and possibly the total electronic momentum.
(The latter case corresponds to electron-electron scattering being dominant,
it obviously does not apply to electron-phonon and electron-impurity scattering
\cite{Conti_1999}.)
Similar arguments allowing for different conserved quantities in 
improved relaxation time approximations
have been put forward in Refs.\citelow{Conti_1999, Roepke_1999, Das_1975, Atwal_2002}.

For simplicity, we focus here on the transverse dielectric function in the
semiclassical (Boltzmann) approximation.
Warren and Ferrell's solution for its version including a carrier lifetime
$\tau$ due to impurity (phonon) scattering is
\begin{equation}
\frac{ \varepsilon_T(q, \omega) }{ \varepsilon_0 } 
= 1 - \frac{ \Omega^2 }{ \omega \varpi }
f_T( z \to 0, u' )
\label{eq:epsT-finite-tau}
\end{equation}
where the complex Lindhard variable is $u' = \varpi / ( q v_F )$,
and the semi-classical limit of the transverse Lindhard function is given by
\begin{equation}
f_T( z \to 0, u ) = 
\frac{ 3 u^2 }{ 2 }
- 
\frac{ 3 u (1 - u^2)}{ 4 } \log \frac{ u - 1 }{ u + 1 }
\,.
\label{eq:def-fT-semicl}
\end{equation}
As in Eqs.\,(\ref{eq:def-fL-q}, \ref{eq:def-fT-q})
above, the branch of the logarithm is taken with a cut along the
negative real axis
($\mathop{\rm Im} \log z \in [-\pi, +\pi]$
and $\mathop{\rm Im} \log z > 0$ if $\mathop{\rm Im} z > 0$).
Note that $\varepsilon_T(q, \omega)$ 
is \emph{not} obtained by adding the imaginary part ${\rm i}/\tau$ to
all occurrences of the frequency $\omega$ 
in Eq.\,(\ref{eq:epsT}).
One factor $1/\omega$ remains exact because it implements the link
between current and polarisation field, ${\bf j}_T = - {\rm i} \omega {\bf P}_T$.

\begin{figure}[htbp]
   \centerline{%
   \includegraphics[height=0.37\textwidth]{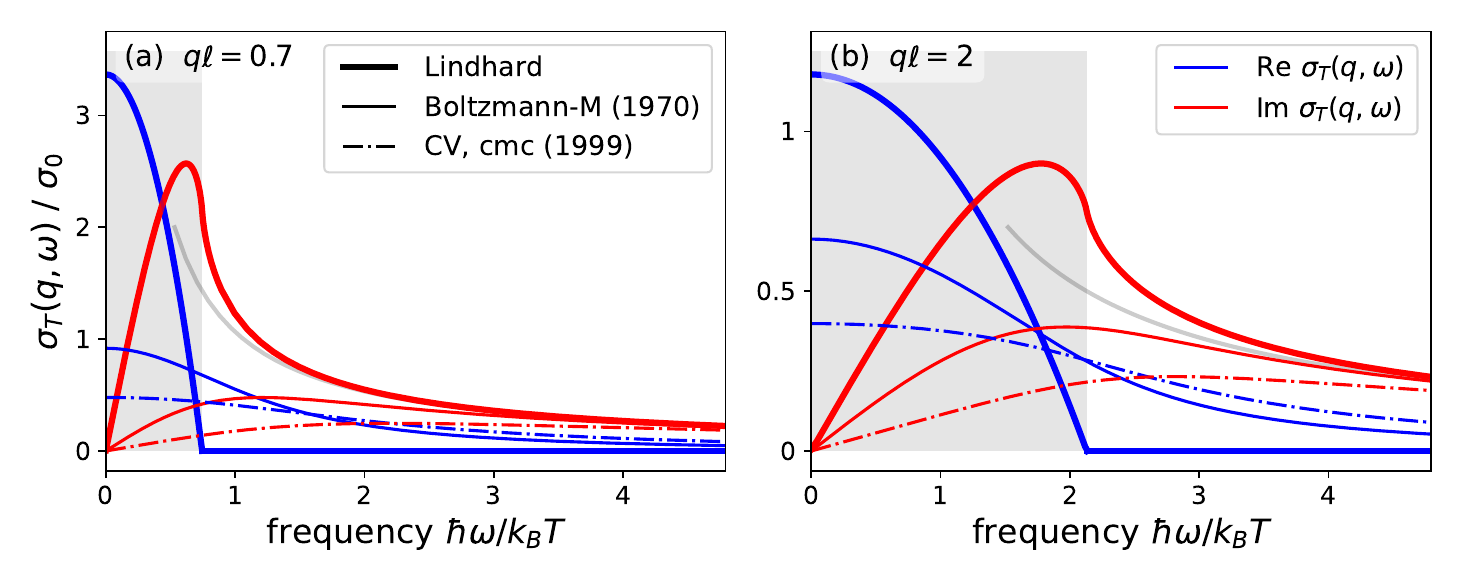}%
   }
   \caption[]{Frequency dependence of nonlocal conductivities 
   for spatial scales comparable to the mean free path $\ell$.
   These data represent the transverse permittivity according
   to  $\varepsilon_T(q, \omega) = \varepsilon_0 + {\rm i} \sigma_T(q, \omega)/\omega$.
   Blue (red) curves: real (imaginary) part of the conductivity.
   In the Lindhard model (thick solid lines), charge carriers do not suffer
   any collisions, but in the shaded frequency band, absorption happens 
   by excitation of electron-hole pairs (Landau damping).
   In the Boltzmann-Mermin \cite{Mermin_1970} (Conti-Vignale \cite{Conti_1999}) 
   models,
   collisions occur at a rate $1/\tau$ and do not conserve (conserve)
   the total momentum of the carriers, respectively.
   The gray line indicates the purely imaginary conductivity
   of the plasma model, $\sigma_{\rm pl}(\omega) = 
   {\rm i} \varepsilon_0 \Omega^2 / \omega$.
   Parameters (typical for gold at room temperature, 
   see Fig.\,\ref{fig:pressure_Mb0}):
   $\ell = v_F \tau \approx 1.58\,c/\Omega$,
   $c / v_F \approx 214$,
   $\hbar/\tau \approx 1.065\,k_B T$. %
   }
   \label{fig:sigmaT-examples}
\end{figure}

The impact of collisions on the nonlocal permittivities
is illustrated in Fig.\,\ref{fig:sigmaT-examples} as a function
of frequency.
The double pole of the collision-free Lindhard model 
[see Eq.\,(\ref{eq:static-eps-LT})] is not visible in this plot, 
as it appears only at very low frequencies, of the order $q v_F (q/k_F)^2$.
We found that a small scattering rate, corresponding to a mean free path of
roughly $3\,{\rm mm}$ (!), is sufficient to suppress this pole.
The sharp features around the border of the Landau damping region 
(gray shaded area) are smoothed out, leading to a significant reduction
of the permittivity. In particular, the DC conductivity is brought 
to a finite value for all values of $q$ and for any reasonable collisional 
model. (The one of Conti and Vignale is defined by 
Eq.\,(\ref{eq:epsT_CV_cmc}) below.)
This implies a static reflection coefficient $r_s(K, 0) = 0$,
as in Eq.\,(\ref{eq:rs-plasma-Drude}) for $\mu = \mu_0$.

We come back to the reflection problem at the zero'th Matsubara frequency
and ask for the fate of the low-frequency permittivity 
of Eq.\,(\ref{eq:static-eps-LT})
when $\tau < \infty$. Taking the limit $\omega \to 0$ in 
Eq.\,(\ref{eq:epsT-finite-tau}) and then expanding for small $q$, 
we find
\begin{equation}
\frac{ \varepsilon_T( q, \omega \to 0 ) 
}{ \varepsilon_0 }
= 
1 + \frac{ {\rm i} \Omega^2 \tau }{ \omega }
\Big(
1 - \frac{ q^2 \ell^2 }{ 5 }
\Big)
\label{eq:epsT-finite-tau-low-omega}
\end{equation}
where $\ell = v_F \tau$ is the mean free path. 
Note that in comparison to Eq.\,(\ref{eq:static-eps-LT}), 
the singularity at $\omega = 0$ has become a first order pole.
The dependence on momentum is quite different and 
involves the mean free path $\ell$, much longer than the Fermi 
wavelength $1 /k_F$.
Equation~(\ref{eq:epsT-finite-tau-low-omega}) gives a transverse
current (compare to Eq.\,(\ref{eq:transverse-current}))
\begin{equation}
{\bf j}_T = {\rm i}\omega \sigma_0 {\bf A} 
- {\rm i} \omega \sigma_0 \frac{ \ell^2 }{ 5 } (\nabla \times {\bf B})
\label{eq:jT-Ohm-plus-diffusive}
\end{equation}
where the first term is Ohm's law with $\sigma_0 = \varepsilon_0 \Omega^2 \tau$,
and the second one accounts for
spatial diffusion on the scale $\ell$, similar to
Chambers' nonlocal conductivity \cite{Warren_1960}.
In the static limit
$\omega \to 0$, however, both terms vanish, and the metal becomes transparent
to magnetic fields, the only transverse fields that survive in this limit.
Hence without solving any interface problem, we know that 
\begin{equation}
\text{lifetime $\tau < \infty$}: \qquad
r_s(K, 0) = 0
\,.
\label{eq:zero-rs}
\end{equation}

As an illustration for the impact of details of scattering processes,
we also provide the result for a collisional transverse permittivity
where the total electronic momentum is conserved \cite{Conti_1999}.
The response function discussed by Conti and Vignale relates the 
current to the vector potential, ${\bf j}_T = \Pi_T {\bf A}_T$. 
It may be understood as a component of the polarization tensor,
$\varepsilon_T - \varepsilon_0 = \Pi_T / \omega^2$,
and connects to the Lindhard function via
$\Pi_T(q, \omega) = - \varepsilon_0 \Omega^2 f_T(z, u)$.

The collisional version of the transverse permittivity
with the above-mentioned assumptions is given by
(in Ref.\citelow{Conti_1999}, atomic units are used, 
and we have restored the required dimensional factors)
\begin{equation}
\frac{ \varepsilon_T(q, \omega) }{ \varepsilon_0 } = 1 
+ \frac{ {\rm i} \Omega^2 \tau }{ \omega }
\frac{ f_T(z, u') }%
{
f_T(z, u')  + 1 - {\rm i} \omega \tau 
}
\,,
\label{eq:epsT_CV_cmc}
\end{equation}
different from Eq.\,(\ref{eq:epsT-finite-tau}), but with the
same $u' = (\omega + {\rm i} / \tau) / (v_F q)$.
The expansion at low frequencies yields
\begin{equation}
\frac{ \varepsilon_T(q, \omega) }{ \varepsilon_0 } = 1 
+ \frac{ {\rm i} \Omega^2 \tau }{ 2 \omega }
\Big(
1 - \frac{ q^2 \ell^2 }{ 10 }
\Big)
\,.
\label{eq:epsT-finite-tau-low-omega-conserved-current}
\end{equation}
This is qualitatively similar to Eq.\,(\ref{eq:epsT-finite-tau-low-omega}),
but has the curious feature that the DC conductivity is divided by $2$.
The simple pole at $\omega = 0$ is sufficient, however, to conclude as 
in Eq.\,(\ref{eq:zero-rs}) 
that static magnetic fields are not reflected.

\begin{figure}[htbp]
   \vspace*{1ex}
   \centering
   \includegraphics[height=0.37\textwidth]{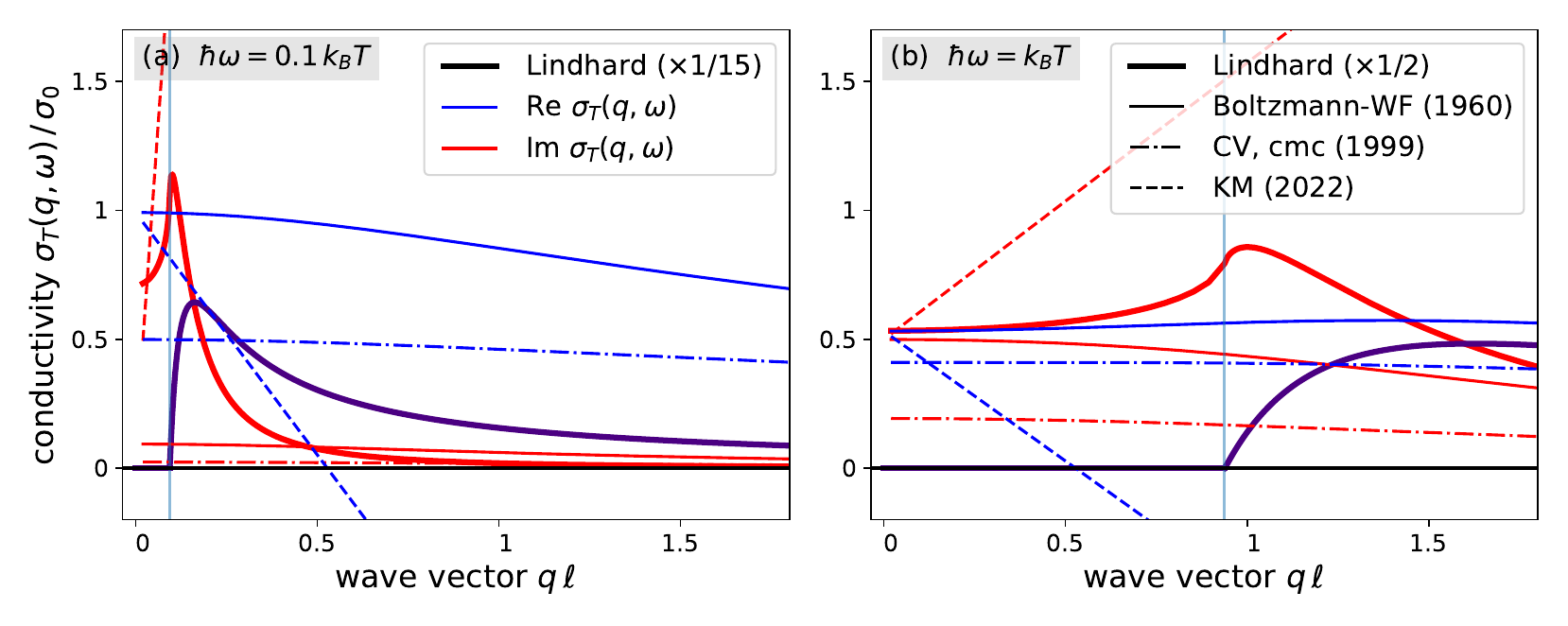}%
   \vspace*{-2ex}
   \caption[]{Nonlocal transverse permittivities vs.\ wave vector $q$, 
   for frequencies lower than and
   comparable to the collision rate $1/\tau$. For convenience,
   we show again the complex conductivity,
   as in Fig.\,\ref{fig:sigmaT-examples}.
   Blue (red) curves: real (imaginary) part of conductivity.
   Thick solid lines: Lindhard model, no collisional broadening
   (data scaled by the factors indicated).
   Solid (dash-dotted) lines: Warren and Ferrell \cite{Warren_1960} 
   based on Eq.\,(\ref{eq:epsT-finite-tau}), 
   Conti and Vignale \cite{Conti_1999} with Eq.\,(\ref{eq:epsT_CV_cmc}),
   see also Fig.\,\ref{fig:sigmaT-examples}.
   Short-dashed lines: Eq.\,(\ref{eq:epsL-KM22a}) proposed by
   Klimchitskaya and Mostepanenko \cite{Klimchitskaya_2022a},
   with parameter $c/v_T \approx 113$.
   The vertical lines give the momentum $q \approx \omega / v_F$
   characteristic for the onset of Landau damping. 
   Other parameters as in Fig.\,\ref{fig:sigmaT-examples}.
   }
   \label{fig:sigmaT_compare}
\end{figure}

The behaviour of the nonlocal permittivities
is illustrated in Fig.\,\ref{fig:sigmaT_compare} as a function
of the wave vector.
The Lindhard results (thick solid lines) have been scaled to fit in the plot, 
illustrating
the significant change brought about by including collisions (solid
and dash-dotted lines). 
In this Figure, the dashed lines show the nonlocal permittivity 
proposed in Ref.\citelow{Klimchitskaya_2022a}
\begin{equation}
\frac{ \varepsilon_T(q, \omega) }{ \varepsilon_0 } = 
1 + \frac{ {\rm i} \Omega^2 \tau }{ \omega (1 + {\rm i} \omega \tau) }
\Big( 1 + {\rm i} \frac{ v_T q }{ \omega } \Big)
\,.
\label{eq:epsL-KM22a}
\end{equation}
where the parameter $v_T$ is of the order of the Fermi velocity.
Its low-frequency limit contains a double pole, so that the reflection
coefficient becomes close to the one of the plasma model. 
The overall behaviour, 
in particular the dependence of Eq.\,(\ref{eq:epsL-KM22a}) on the wave vector $q$, 
is quite different from the results obtained by Lindhard, however.
This occurs even for wave vectors much smaller than $1/\ell$
that are relevant for distances larger than $\approx 50\,{\rm nm}$.
In addition, the model shows negative damping (\emph{cf.}\ the real part 
of the conductivity) for $q > v_T \tau$.

\section{Conclusion}

We have computed the reflection amplitudes $r_p, r_s$ for the
zero'th term in the Matsubara representation of the Casimir pressure
between conducting plates. 
The starting point is the low-frequency limit of the Lindhard 
dielectric functions, evaluated from first principles,
and the matching of the relevant electromagnetic fields 
at the metal-vacuum interface. 
The results confirm the predictions of the Drude model. 
The thermal anomaly remains open and calls for an understanding
why the Drude predictions deviate from the experimental Casimir
pressure data.

\paragraph{Acknowledgments.}
I thank the participants of the 2024 International Casimir Symposium 
in Piran (Slovenia) for inspiring questions and discussions. 
This research was funded by the Deutsche Forschungsgemeinschaft 
(DFG, German Research Foundation) within SFB 1636, ID 510943930,
Projects No.\ A01 and A04).

\appendix

\section{Spatial dispersion and magnetic response}
\label{a:spatial-dispersion}

In a bulk system, longitudinal fields are irrotational and can be
written as gradients. Transverse fields are divergence-free and
can be expressed via the vector potential in the Coulomb gauge $\nabla \cdot
{\bf A} = 0$. 
The polarization response of a bulk medium to the two types of fields
is 
\begin{equation}
{\bf P} = \varepsilon_L {\bf E}_L 
+ \varepsilon_T {\bf E}_T
- \varepsilon_0 {\bf E}
\label{eq:}
\end{equation}
whose longitudinal part determines the charge density 
($\rho = - \nabla \cdot {\bf P}$).
The first term can be written as a response to the total 
field ${\bf E} = {\bf E}_L + {\bf E}_T$
\begin{equation}
{\bf P} = (\varepsilon_L - \varepsilon_0){\bf E} 
+ (\varepsilon_T - \varepsilon_L){\bf E}_T
\,.
\label{eq:}
\end{equation}
The second term which is transverse, gives rise to a magnetisation (current):
\begin{equation}
\partial_t {\bf P}_T = \nabla \times {\bf M}
\qquad \text{or} \qquad
(\varepsilon_T - \varepsilon_L) {\bf E}_T = 
- \frac{ 1 }{ \omega } {\bf q} \times {\bf M}
\label{eq:}
\end{equation}
where the second form is written in Fourier space. 
Apply ${\bf q} \times {}$ and use the Faraday equation to eliminate
${\bf q} \times {\bf E}_T$ 
\begin{equation}
(\varepsilon_T - \varepsilon_L) {\bf B} = 
\frac{ q^2 }{ \omega^2 } {\bf M}
\,.
\label{eq:}
\end{equation}
Hence the magnetic susceptibility, defined by the linear
response ${\bf M} = \chi {\bf H}$ 
can be identified as [Eq.(1.6) of \citelow{Lindhard_1954}]
\begin{equation}
\varepsilon_T - \varepsilon_L
= 
\frac{ q^2 }{ \omega^2 } \frac{ \chi }{ \mu } 
=
\frac{q^2}{\omega^2}
\Big( \frac{ 1 }{ \mu_0 } - \frac{ 1 }{ \mu } \Big)
\label{eq:T-L-and-mu}
\end{equation}
where $\mu$ is the permeability.
In the second expression, we have used 
the conventional form of the magnetic flux density:
${\bf B} = \mu_0 ({\bf H} + {\bf M}) = \mu {\bf H}$. 
Therefore
\begin{equation}
\mu = \mu_0 ( 1 + \chi )
\qquad\text{or}\qquad
\chi = %
\frac{ \mu }{ \mu_0 } - 1
\,.
\label{eq:}
\end{equation}
In the following, the long-wavelength limit $q \to 0$ is of particular
interest. If $\varepsilon_L$ has a finite limit, it plays the role
of the local dielectric function.
The difference between longitudinal and transverse fields
is then interpreted as an effective magnetic susceptibility.
This is how the Landau diamagnetism is recovered from the Lindhard
dielectric functions (see Sec.\,\ref{s:s}).

\section{Lindhard functions}
\label{a:Lindhard-functions}

\begin{figure}[bht]
\includegraphics*[height=0.45\textwidth]{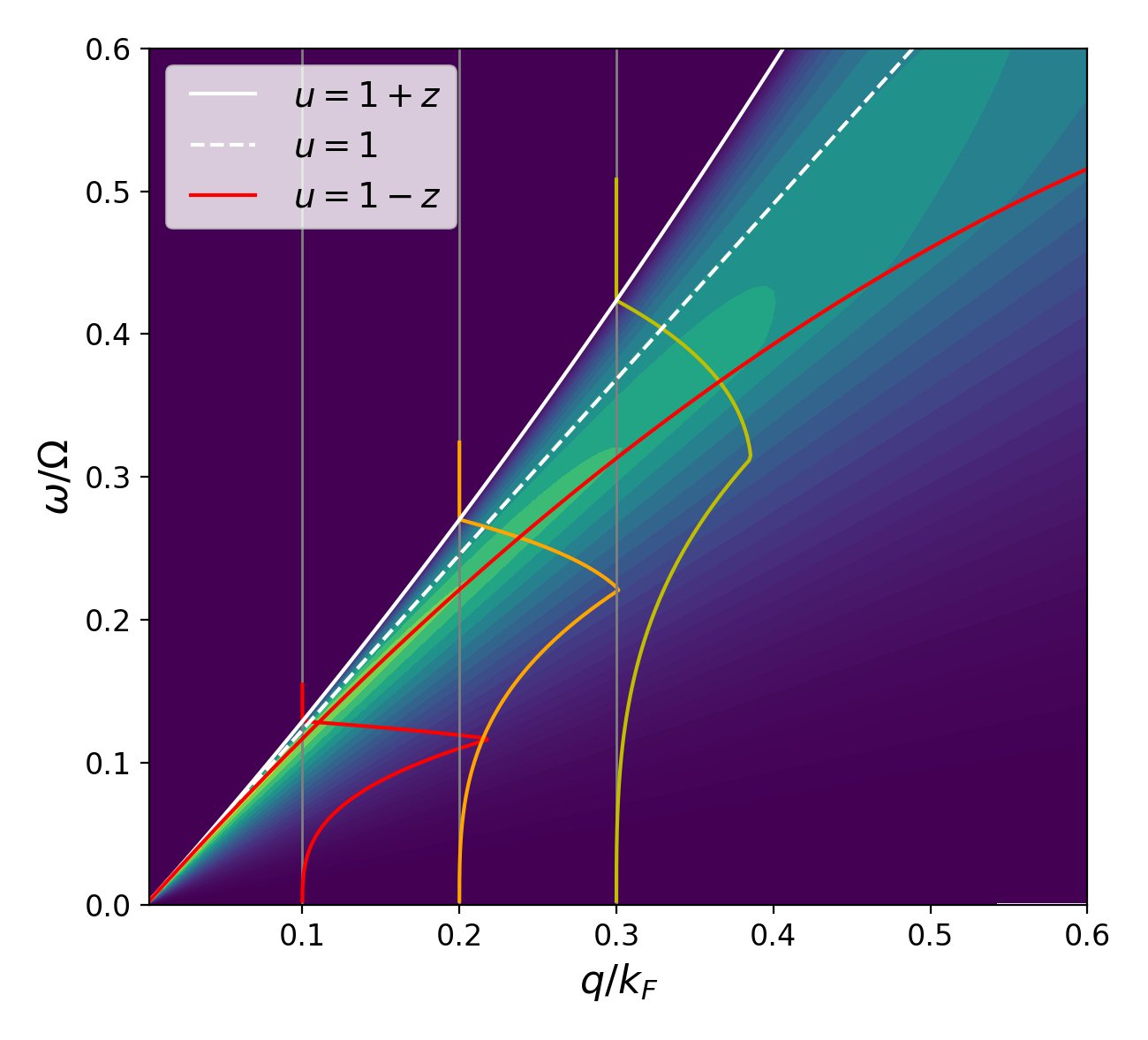}%
\hspace*{3mm}%
\includegraphics*[height=0.45\textwidth]{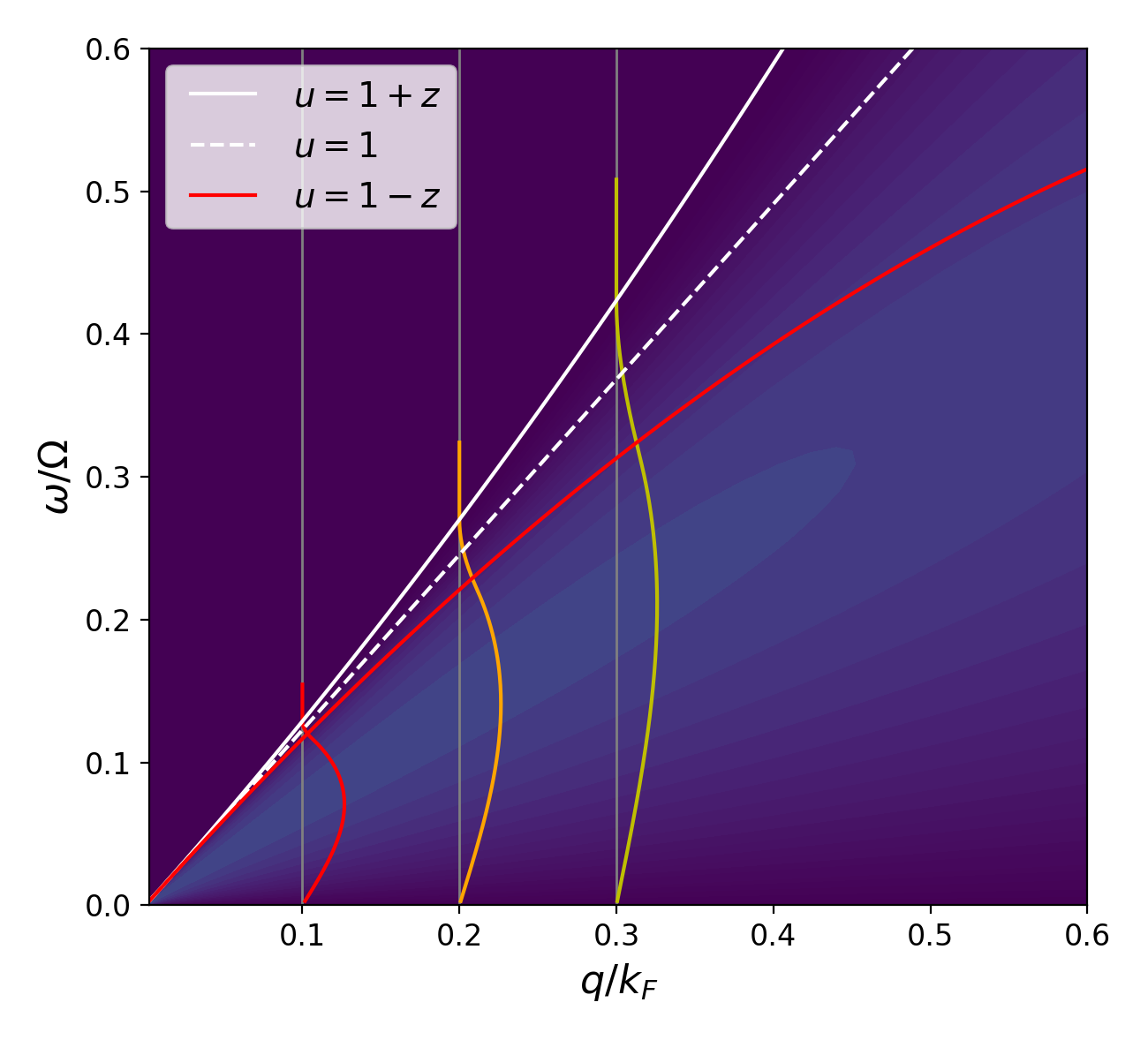}%
\caption[]{Imaginary part of Lindhard dielectric functions in 
the $q\omega$-plane; (left) the longitudinal, (right) the transverse
version. 
For better visibility, the data have been multiplied by $\omega$
(so that actually the real part of the nonlocal conductivities is plotted).
The white solid line corresponds to those parameter combinations
of Lindhard's $u = \omega / (q v_F)$ and $z = q / (2 k_F)$ variables
(as given in the legend) 
that delimit the regions where particle-hole excitations are 
kinematically allowed. 
The color code uses the same maximum value for both plots. 
The vertical colored lines give a cut along the frequency axis 
for three fixed momenta (same scaling in the two plots). 
As one crosses the red line, kinks appear.
Parameter: plasma frequency $\Omega \approx 0.81\,k_F v_F$,
typical for the valence electron density in gold. 
No scattering losses included; realistic lifetimes 
($\Omega \tau \gtrsim 100$)
would make no significant changes on this scale.
}
\label{fig:absorption}
\end{figure}

The real and imaginary parts of Eq.\,(\ref{eq:epsL-Lindhard}) can be
found by identifying the branch points of the logarithms. 
The result for real $u$ is
\begin{eqnarray}
\mathop{\rm Re} f_L( z, u ) &=&
\frac12 
+
\frac{ 1 - (z - u)^2 }{ 8 z }
\log \left|\frac{ z - u + 1 }{ z - u - 1 }\right|
\nonumber\\
&&
\hphantom{\frac12 
}
{} 
+
\frac{ 1 - (z + u)^2 }{ 8 z }
\log \left|\frac{ z + u + 1 }{ z + u - 1 }\right|
\,,
\\
\mathop{\rm Im} f_L( z, u ) &=&
\begin{cases}
\displaystyle
\frac{\pi u}{2} & \mbox{for } u + z < 1
\\[2ex]
\displaystyle
\frac{\pi}{8 z}[1 - (z - u)^2] & \mbox{for } |u - z| < 1 < u + z
\\[2ex]
\displaystyle
0 & \mbox{for } 1 < |u - z|
\,.
\end{cases}
\label{eq:}
\end{eqnarray}
The corresponding regions of absorption (imaginary part $\varepsilon_L$)
are marked in Fig.\,\ref{fig:absorption}(left) in the $q\omega$-plane.
They are mostly concentrated around the region $\omega \approx q v_F$
(dashed line).
The physical process behind is the excitation of electron-hole pairs 
around the Fermi edge, also known as Landau damping.

The real and imaginary parts of the transverse 
permittivity~(\ref{eq:epsT}) are
\begin{align}
\mathop{\rm Re} f_T(z, u) &=
\frac{ 3 }{ 8 }( 1 + 3 u^2 + z^2 )
\nonumber\\
& \phantom{=
}
- 
\frac{ 3 [1 - (z-u)^2]^2}{ 32 z} \log \left| \frac{ z - u + 1 }{ z - u - 1 } \right|
- 
\frac{ 3 [1 - (z+u)^2]^2}{ 32 z} \log \left| \frac{ z + u + 1 }{ z + u - 1 } \right|
\,,
\nonumber\\
\mathop{\rm Im} f_T(z, u) &=
\begin{cases}
\displaystyle
- \frac{3 \pi u}{4}(1 - u^2 - z^2)
& \mbox{for } u + z < 1
\\[2ex]
\displaystyle
- \frac{3\pi}{32 z}[1 - (u - z)^2] & \mbox{for } |u - z| < 1 < u + z
\\[2ex]
\displaystyle
0 & \mbox{for } 1 < |u - z|
\,.
\end{cases}
\label{eq:}
\end{align}
For the sign of the imaginary part which differs from Lindhard's 
paper \cite{Lindhard_1954}, see Refs.\citelow{DresselGruener_book, Levine_2008}.
The plot in Fig.\,\ref{fig:absorption} shows by comparison that
the transverse absorption is somewhat smaller and more smoothed out
towards lower frequencies.
For complex values of $u$, we evaluate the logarithms in 
Eqs.\,(\ref{eq:epsL-Lindhard}, \ref{eq:epsT}) directly.

%

\end{document}